\documentclass[structabstract]{aa}  
\usepackage{graphicx}
\usepackage{txfonts}
\usepackage{natbib}

\usepackage{dsfont}
\bibpunct{(}{)}{;}{a}{}{,}
\begin{document}
   \title{On the velocity field of sunspot penumbrae}
   
   \subtitle{I. A global view}

   \author{M. Franz
          \and
          R. Schlichenmaier}

   \institute{Kiepenheuer Institut f\"ur Sonnenphysik,
              Sch\"oneckstra\ss{}e 6, D-79104 Freiburg\\
              \email{morten@kis.uni-freiburg.de}}

   \date{Submitted:  06 August 2009, Accepted: 24 September 2009}

% \abstract{}{}{}{}{} 
% 5 {} token are mandatory
 
\abstract
  % context heading (optional), leave it empty if necessary  
  {}
{We investigated the vertical penumbral plasma flow on small spatial scales using data recorded by the spectropolarimeter of the solar optical telescope onboard {\it{Hinode}}.}
{We computed maps of apparent Doppler velocities by comparing the spectral position of the Fe I 630.15 nm \& Fe I 630.25 nm lines with the averaged line profiles of the quiet Sun. To visualize the flow pattern in the low photosphere, we used a bisector of the wing of the absorption lines. Due to the small heliocentric angle ($3^{\circ} \le \Theta \le 9^{\circ}$) of our data sets, the horizontal component of the Evershed flow (EF) does not contribute significantly to the line shift.}
{We found that in the quiet Sun (QS), the area showing up-flows is always larger than the one exhibiting down-flows. In the penumbra, up-flows dominate only at low velocities $|\rm{{v}}_{\rm{dop}}| \le 0.4$ km s$^{-1}$, while at larger velocities $|\rm{{v}}_{\rm{dop}}| \ge 0.6$ km s$^{-1}$ down-flows prevail. Additionally, the maximal up-flow velocity in penumbrae is smaller, while the maximal down-flow velocity is larger with respect to the QS velocities. Furthermore, on a spatial average, the penumbra shows a red-shift, corresponding to a down-flow of more than 0.1 km s$^{-1}$. Up-flows are elongated and appear predominately in the inner penumbra. Strong down-flows with velocities of up to 9 km s$^{-1}$ are concentrated at the penumbra-QS boundary. They are magnetized and are rather round in shape. The inner penumbra shows an average up-flow, which turns into a mean down-flow in the outer penumbra. The up-flow patches in the inner penumbra and the down-flow locations in the outer penumbra could be interpreted as the sources and the sinks of the EF. We did not find any indication of roll-type convection within penumbral filaments.}
  {}

   \keywords{Sunspots - Sun: photosphere}

   \maketitle
%
%______________________________________________________________

\section{Introduction}
Although penumbrae of sunspots have been subject to scientific investigation for a long time, there remain a lot of open questions regarding  e.g. their filamentary fine structure, the energy transport and the plasma dynamics on scales of less than 1". 

The {\it{gappy}} model \citep{Spruit2006} assumes the existence of field-free, radially aligned gaps below the $\tau=1$ level intruding into a potential field above. Dark cores of penumbral filaments \citep{Scharmer2002} are explained as a variation of the $\tau=1$ level across the tip of the field-free gap, which is similar to  the explanation of dark lanes in umbral dots given by \citet{Schuessler2006}. Even though the {\it{gappy}} model provides an explanation not only for the dark cores but also for the brightness of the penumbra, it poses a problem as it does not explain the magnetized Evershed flow (EF) \citep{Rezaei2006}. 

In the {\it{uncombed}} model \citep{Solanki1993a}, nearly horizontal flux tubes are assumed to be embedded in a less inclined magnetic background field \citep{Schlichenmaier1998a,Schlichenmaier1998b}. Strong gradients of atmospheric parameters are encountered for any line of sight (LOS) penetrating both the background atmosphere and the flux tube. It has recently been shown that dark cores of penumbral filaments can be explained within the framework of this model by taking the hot EF, which is embedded in a stratified atmosphere, into account \citep{RuizCobo2008}. Despite the success of the {\it{uncombed}} model in explaining e.g. the EF as well as the existence of bright filaments and their migration, the permanent stability of magnetic flux tubes in the penumbral photosphere is questionable.

Recently, significant progress regarding simulations of entire sunspots has been made. Different 3D MHD codes were used to model sunspots in slab geometry \citep{Heinemann2007} or even the entire spot at once \citep{Rempel2009}, which reproduces a variety of observed features, e.g. dark cored penumbral filaments and their migration. In order to either endorse or disprove these results and to gain a deeper understanding of the convective nature of the energy transport in the inclined magnetic field of the penumbra, it is thus crucial to investigate the plasma flow in the low photosphere on scales of less then 1").

So far velocities are either inferred from inversion (e.g. SIR, Milne-Eddington) \citep{BellotRubio2007, Scharmer2008} or magnetograms that are constructed in the far wings of magnetic absorption lines, thereby identifying strongly blue- or red-shifted polarization signals \citep{Ichimoto2007}. The strength of the signal in the far wing, however, does not only depend on the velocity itself, but also on field inclination or gas temperature \citep{BellotRubio2009}.

In this contribution, we present measurements of line shifts in the wings of the Stokes I profile, which enables us to analyze the Doppler velocities, directly and quantitatively.

%______________________________________________________________

\section{Observation}
We used data obtained by the spectropolarimeter (SP) \citep{Lites2001} of the solar optical telescope (SOT) \citep{Tsuneta2008} onboard the {\it{Hinode}} satellite. The SP records the Stokes spectra of the two iron lines at 630.15 nm and 630.25 nm, Land\'e factors of $\rm{g}=1.67$ and $\rm{g}=2.5$, respectively. By scanning the spectrograph slit across the target in steps of 0."15 and achieving a wavelength sampling of 2.15 pm/pixel, two dimensional maps of the field of view (FOV) are obtained. Since the width of the slit is equivalent to 0."16 and the pixel size along the slit is 0."16 on average, normal SP scans provide a spatial resolution of 0."32 \citep{Centeno2009}. For an exposure time of 4.8 s per slit position, the 1$\sigma$ noise level in the Stokes spectra is of the order of $10^{-3}{\cdot}\rm{I}_{\rm{c}}$, where ${\rm{I}}_{\rm{c}}$ is the continuum intensity. 

\begin{table}[h]
\begin{center}
	\caption{Data used in this study. From left to right: Sample name, NOAA active region, date of observation, heliocentric angle of penumbra, or field of view in quiet Sun samples.}
	\begin{tabular}{cccc}
		\multicolumn{4}{c}{Data Sample}\\		
		\hline
		\hline
		\\[-2ex]
		{Name}&{NOAA} & {Observation} & {$\mu$}\\
		\hline
		Spot A & {10923} & {Nov 14$^{\rm{}th}$ 2006} & {0.982 - 0.996}\\
		Spot B & {10923} & {Nov 14$^{\rm{}th}$ 2006} & {0.980 - 0.994}\\
		Spot C & {10930} & {Dec 11$^{\rm{}th}$ 2006} & {0.991 - 0.999}\\
		Spot D & {10933} & {Jan 05$^{\rm{}th}$ 2007} &{0.996 - 1.000}\\
		QS 1 & {...} & {Mar 10$^{\rm{}th}$ 2007} & {0.970 - 1.000}\\
		QS 2 & {...} & {Sep 06$^{\rm{}th}$ 2007} & {0.988 - 1.000}\\
		\hline
	\end{tabular}
	\label{Tab_1}
\end{center}
\end{table}

For our study, we obtained the raw data summarized in Table \ref{Tab_1} and reduced it, using the IDL routine 'sp\_prep.pro' provided with the 'Solar Soft' package. From the Stokes spectra, we computed maps of continuum intensity, $\rm{I}_{\rm{c}}$, total polarization $\rm{P}_{\rm{tot}}=\int [(\rm{Q}^{2}+\rm{U}^{2}+\rm{V}^{2})/\rm{I}_{\rm{c}}^{2}]^{1/2}\rm{d\lambda}$ and Doppler velocity along the LOS, ${\rm{v}}_{\rm{dop}}$. Furthermore, we increased the contrast of the maps of $\rm{I}_{\rm{c}}$ and $\rm{P}_{\rm{tot}}$ by applying a standard IDL high-pass filter to the pictures and then convolved the result with the original. As a consequence, the picture appears sharper, and it becomes easier to distinguish between bright and dark features in the penumbra (e.g. the lateral brightening and the dark core of a filament). While the calculation of the maps of $\rm{I}_{\rm{c}}$ and $\rm{P}_{\rm{tot}}$ is straightforward, the derivation of ${\rm{v}}_{\rm{dop}}$ requires some additional explanation.

\section{Method}
\label{Method}

Since {\it{Hinode}} spectra lack telluric lines, it is not possible to use the latter to calibrate the velocity scale absolutely. To overcome this problem, we used an average QS profile for calibration purposes. The disadvantage of this method is, however, that the respective lines are displaced from laboratory wavelength by several effects, including the convective blue-shift (CBS) \citep{Dravins1982}. \citet{Beck2005} used the SIR code with a two-component model atmosphere of \citet{Borrero2002} to determine the CBS at disk center. The corresponding velocities of the line core of the average QS profile of Fe I 630.15 and Fe I 630.25 were calculated to be $\rm{-185}$ m s$^{-1}$ and $\rm{-262}$ m s$^{-1}$ respectively (negative velocity corresponds to blue-shift).

\subsection{Velocity Calibration Using Quiet Sun Profiles}

We have argued that the line core of an average Stokes I profile of the QS is suitable for an absolute wavelength calibration, if it is corrected for the CBS. However, two effects, namely the moat flow and the influence of the magnetic field, have to be considered in the surroundings of sunspots.

Due to the residual heliocentric angle, the radially outward oriented moat flow \citep{Vargas2008} would lead to a systematic offset in the calibration, if only the Stokes I profiles of one side of the spot vicinity were taken into account. Additionally, a magnetic field lifts the degeneracy of the atomic levels, which causes the profile to split and makes it difficult to determine the position of the line core correctly \citep{Solanki1993b}.

\begin{figure}[h]
	\centering
		\includegraphics[width=\columnwidth]{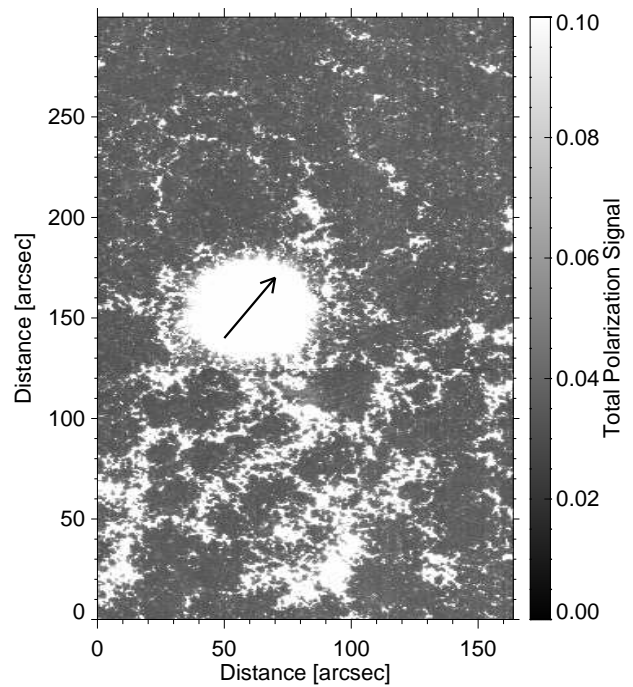}
		\caption{Total polarization of data sample Spot D. The spectra of the {pixels} marked in white were not considered for an averaged Stokes I profile of the QS. The arrow points towards disk center at $\rm{(x,y)} \sim \rm{(87",195")}$.}
		\label{Franz_fig00}
\end{figure}

Therefore, we decided to average all profiles of the QS surrounding the sunspot, but to exclude those profiles that exhibit a split. In order to meet this requirements, we created a mask which omits the entire spot and all pixels with significant magnetic field strength. For these purposes, we used $\rm{P}_{\rm{tot}}$ as a proxy and defined a cut-off criterion $\rm{P}_{\rm{tot}} > 0.1 \cdot max(\rm{P}_{\rm{tot}})$ for all pixels whose Stokes I spectra should not be averaged. As an example, the mask of Spot D is indicated in white in Fig. \ref{Franz_fig00}.

Finally, the line minima of Fe I 630.15 and Fe I 630.25 in the average Stokes I profile of the QS are determined by fitting a 2$^{\rm{nd}}$ order polynomial to the core of the respective line (cf. dotted line in Fig. \ref{Franz_fig01}).

\subsection{Velocity Calibration Using Umbral Profiles}

\citet{Beckers1977} showed that there is no vertical plasma motion in umbrae of sunspots at disk center. Consequently, an average umbral profile should be red-shifted when compared to the line core of an average QS profile. To substantiate the values given by \citet{Beck2005}, we followed a procedure described in \citet{Rezaei2006}. 

Due to the strength of the magnetic field, the core of umbral Stokes I profiles splits, and it is difficult to determine the core position correctly. Therefore, we determined the center-position - midways between the two lobes -  of umbral Stokes V profiles, which corresponds to the line core of the respective Stokes I profile. This parameter can be calculated with reasonable precision if molecular blends are not too strong and both lobes are well defined. If additionally LTE can be assumed, and no velocity gradients are present in the line forming region, e.g. above the umbra, Stokes V profiles are strictly antisymmetric with respect to the center-position \citep{Auer1978}. Thus, the center-position of antisymmetric Stokes V profiles defines a frame of rest on the solar surface.

As a measure for the degree of antisymmetry, we used the amplitude asymmetry. It is defined as the difference of the absolute amplitudes of the two lobes like
\begin{equation}
\delta \rm{a} = \frac{\rm{a}_{\rm{r}} - \rm{a}_{\rm{b}}}{\rm{a}_{\rm{r}} + \rm{a}_{\rm{b}}},
\end{equation}
with $\rm{a}_{\rm{r}}$ and $\rm{a}_{\rm{b}}$ being the amplitude of the red and the blue lobe, respectively. We used a parabola fit around the extrema of both lobes to determine its amplitude.

Following the arguments above, we considered only umbral Stokes V profiles with a $|\delta \rm{a} |< 0.01$ and determined their center position. The results are summarized in Table \ref{Tab_2}.

\begin{table}[h]
\begin{center}
	\caption{From left to right: Data sample, CBS of the average QS if the umbra is assumed to be at rest, and RMS values the center position of umbral Stokes V profiles with $|\delta \rm{a} |< 0.01$.}
	\begin{tabular}{ccccc}
	\multicolumn{5}{c}{Wavelength Calibration}\\
		\hline
		\hline
		\\[-2ex]
			{Name} & \multicolumn{2}{c}{CBS [m s$^{-1}$]} & \multicolumn{2}{c}{RMS Center Position [m s$^{-1}$]}\\
			{} & {Fe I 630.15} & {Fe I 630.25} & {Fe I 630.15} & {Fe I 630.25}\\
		\hline
		{Spot A} & {-231}& {-296} & {93} & {83}\\
		{Spot B} & {-245}& {-315} & {94} & {82}\\
		{Spot C} & {-201}& {-295} & {100} & {92}\\
		{Spot D} & {-231}& {-344} &  {147} & {107}\\
		\hline
	\end{tabular}
		\label{Tab_2}
\end{center}
\end{table}

We find that in all our data samples the average QS is indeed blue-shifted with respect to the average center position of umbral Stokes V profiles obeying $|\delta \rm{a} |< 0.01$. Within the listed RMS errors (cf. Tab. \ref{Tab_2}), the values for the CBS agree with the ones reported by \citet{Beck2005}. Furthermore, we can confirm the applicability of the procedure reported by \citet{Rezaei2006}.

\subsection{Flow velocity in the deep photosphere}

\begin{figure}[h]
	\centering
		\includegraphics[width=\columnwidth]{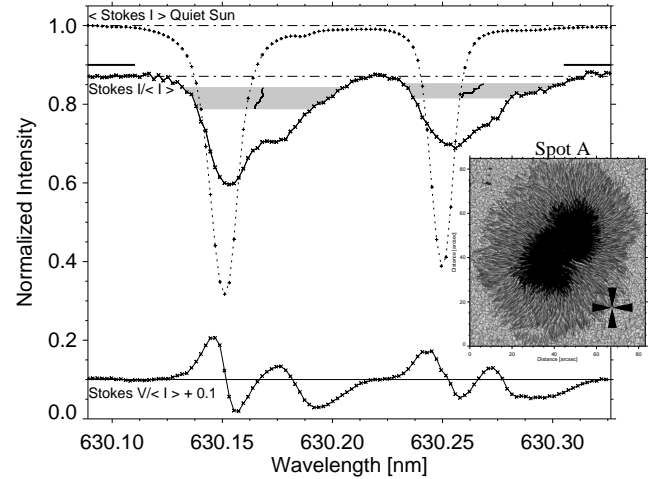}
		\caption{ Stokes I and Stokes V profiles (solid line, measurements plotted with 'X' symbols) from a down-flow region (marked by the cross hairs in the inlet) in Spot A as well as the average Stokes I profile from the QS (dotted line, measurements indicated by '+' symbols). The mean value of the bisector (solid) of the line wing (shaded gray) is used to derive the Doppler-shift of that pixel. $\rm{I}_{\rm{c}}$ (dashed-dotted) is calculated by averaging the measurements below the two bold lines on the left and right sides of the spectrum.}
		\label{Franz_fig01}
\end{figure}

As we are interested in the plasma flow in the low photosphere, it is necessary to compute the line position as close to the continuum as possible. On the one hand, the line forming region of Fe I 630.15 encompasses higher atmospheric layers if compared to Fe I 630.25, but on the other hand, due to the lower Land\'e g factor, the former is less sensitive to disturbances by magnetic fields and instrumental crosstalk. In Fe I 630.15, a blend in the red wing close to the continuum obscures the profile and makes it impossible to derive a reliable bisector above $0.9 \cdot \rm{I}_{\rm{c}}$ (cf. dotted line in Fig. \ref{Franz_fig01}). Similar obstacles occur in the spectra of Fe I 630.25 originating from the innermost penumbra. In this area, the magnetic field, and thus the line broadening, is so strong that the adjacent iron line serves as a blend, leaving the blue part of the continuum not defined above $0.9 \cdot \rm{I}_{\rm{c}}$. Therefore, we restricted the determination of the bisector of each line to a band of $0.7 \le \rm{I}_{\rm{c}} \le 0.9$ (gray shaded region in Fig. \ref{Franz_fig01}).

Next, we determined the bisector of both lines with sub-pixel accuracy by interpolating linearly between the spectral measurements. From that bisector (solid line in the middle of the shaded area in Fig. \ref{Franz_fig01}), we obtained an average spectral position of the line wing. Any offset between the latter position and the position of the line core of the averaged QS profile is interpreted as a Doppler-shift caused by the velocities of the plasma. Finally, the Doppler maps (cf. Fig. \ref{Franz_fig04}) are obtained by computing the velocities that correspond to the Doppler-shifts at any given spatial position. 

As mentioned above, the low photospheric temperature of the umbra allows the formation of molecules. Their absorption bands mask the two iron lines to an extent where it becomes unreliable to compute a bisector. Therefore, the Doppler velocity of pixels with an $\rm{I}_{\rm{c}}$ of less that 33 \% of that of the average QS were manually set to zero (cf. Fig. \ref{Franz_fig04}). 

\subsection{Error discussion}

The biggest challenge in the calibration process is the determination of the line core position of the average QS profile. Due to the finite spectral resolution, we have seven spectral measurements to fit the core. Using, for example, only five measurement to fit the core in sample 'Spot A' changes the CBS of the QS by $+$40 m s$^{-1}$. In the QS significant magnetic fields are predominately found in the intergranular down-flow regions. As a result, we exclude more pixels showing down-flows and introduce a bias to our calibration (cf. Fig. \ref{Franz_fig00}). This bias leads to an overall blue-shift of the maps of the order of $-10$ m s$^{-1}$ to $-30$ m s$^{-1}$ in all data samples. The thermal drift in spectral direction is corrected by the 'sp\_prep.pro' calibration routine. In all our data samples, it is within the range of $\pm$50 m s$^{-1}$. Furthermore, we have shown that the values of the CBS of the QS inferred from model calculations and the values obtained from measurements differ by not more than $\pm$80 m s$^{-1}$.

Due to the difficulties mentioned above (especially the uncertainties in the position of the line core of the average QS) and since our fit procedure allows us to determine the line core within an accuracy of one tenth of a pixel, we assume a precision of $\pm0.1$ km s$^{-1}$ for our wavelength calibration. As we do not consider any velocity signal $-0.1\,\rm{km\,s}^{-1} \le \rm{v_{dop}} \le 0.1\,\rm{km\,s}^{-1}$ in the following, we are confident that any up- or down-flow is an actual velocity signal.

\section{Results}

\subsection{Differences between Fe I 630.15 and Fe I 630.25}

The resulting velocities derived from the Doppler-shifts of the two line wings are comparable: The RMS values of the maps deviate less than $0.17$ km s$^{-1}$. In general, the maps deduced from the wing of Fe I 630.15 show an overall blue-shift with respect to the maps deduced from the wing of Fe I 630.25.

It seems that in all cases with significant differences ($|\rm{v_{dop\:630.15}}-\rm{v_{dop\:630.25}}| > 0.5$ km s$^{-1}$), the Fe I 630.25 line exhibits Doppler-shifts closer to the continuum than the Fe I 630.15 line. This behavior is shown for a selected pixel with a large velocity discrepancy ($\Delta\rm{v}_{\rm{dop}}=1.5$ km s$^{-1}$) in Fig. \ref{Franz_fig01}. Here the Stokes I and Stokes V profiles from a down-flow region in the outer penumbra of Spot A are depicted. The average Stokes I profile from the QS is indicated by the dotted line.

We see a significant difference in the asymmetry of both lines, especially close to the continuum, which has a strong impact on our calculated velocity. In this example, the red wing of Fe I 630.15 is strongly shifted above $0.35 \cdot \rm{I}_{\rm{c}}$. The respective wing of Fe I 630.25, however, shows a strong red-shift only above $0.77 \cdot \rm{I}_{\rm{c}}$. Thus, the resulting velocities show significant differences. While the Doppler-shift of the average line wing ($0.7 \le \rm{I}_{\rm{c}} \le 0.9$) of Fe I 630.15 corresponds to 7.8 km s$^{-1}$, the respective Doppler-shift of Fe I 630.25 leads to 6.3 km s$^{-1}$, that is difference of 1.5 km s$^{-1}$. In some cases, the asymmetries are above $0.9 \cdot \rm{I}_{\rm{c}}$ and remain undetected in Fe I 630.25. Note, however, that only the amplitude of the velocities, but neither the morphology of up- and down-flow patterns nor the direction of the flow are affected by this deviations. 

We do not have an explanation for these differences, but we are tempted to ascribe the variations in the maps of $\rm{v_{dop}}$ to differences of the respective line parameters. Both lines have different g-factors, with Fe I 630.25 being more sensitive to magnetic fields. The line core of Fe I 630.15 forms at $\rm{log}\,\tau=-2.9$, while the core of Fe I 630.25 forms around $\rm{log}\,\tau=-2.0$ and therefore encompasses lower atmospheric layers. As the strength of the magnetic field decreases with hight, its influence on the Fe I 630.25 line is even more significant \citep{Khomenko2007}. 

\subsection{Global Velocity Field}

As all data is taken almost at disk center, the maps of Doppler velocities (see Fig. \ref{Franz_fig04} as an example) are dominated by plasma flows normal to the surface. We attribute any asymmetry between the center-side penumbra, which is overall blue-shifted, and the limb side, which shows an overall red-shift, to a combination of the horizontal EF and projection effects caused by the small heliocentric angle of the respective data sets. The up-flow in Spot D appears predominately but not exclusively in the inner penumbra. It encompasses peak velocities around $-$2.0 km s$^{-1}$. The down-flow shows peak velocities of up to 5.0 km s$^{-1}$, which are generally located at the penumbra-QS boundary, while weaker down-flows ($0\,\rm{km\,s}^{-1} \le \rm{v_{dop}} \le 2\,\rm{km\,s}^{-1}$) appear at all radial distances. The maximal up-flow velocities found in the other samples are very similar. The maximum down-flow velocity, however, shows variation of up to 80 \% (cf. Table \ref{Tab_3}). Interestingly, the former seems to change on a temporal scale of hours. In Spot A it peaks around 9.0 km s$^{-1}$, 9 hours later in Spot B it peaks at 6.9 km s$^{-1}$. Note that on a spatial average the penumbra shows a significant red-shift corresponding to the down-flow velocity listed in Table \ref{Tab_3}.

\begin{figure}[h]
	\centering
		\includegraphics[width={\columnwidth}]{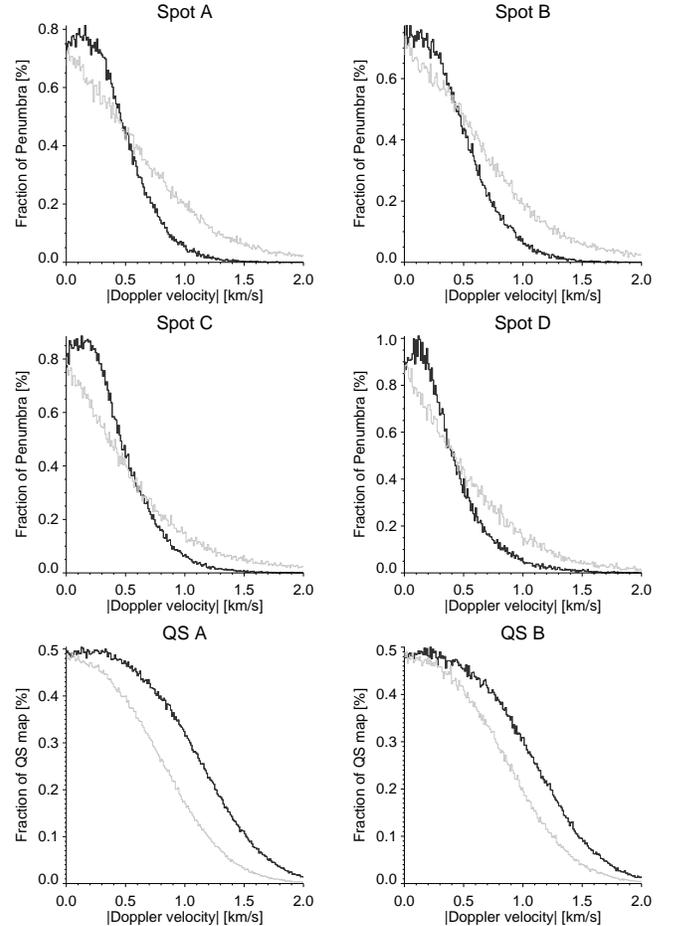}
		\caption{Histogram of binned Doppler velocities in sunspot (top rows) and QS (bottom row) samples in which the fraction of the penumbra or QS is depicted vs. the absolute value of the Doppler velocity. Black indicates up-flow, gray is down-flow. In all these measurements, the error is $\pm$0.1 km s$^{-1}$ and indicated by the black error bars.}
 		\label{Franz_fig02}
\end{figure}

The histograms in Figure \ref{Franz_fig02} show the distribution of velocities in the data sets in more detail. The absolute value of a velocity range of 10 m s$^{-1}$ is binned into one measurement and plotted on the abscissae. The fraction of the penumbra showing the respective velocity range is depicted on the ordinate. Down-flows appear in gray, while up-flows are drawn in black. Keep in mind that no unambiguous flow direction can be derived for $|\rm{v}_{\rm{dop}}| < 0.1 \rm{km\,s}^{-1}$. For reasons of comparison, QS data is evaluated in the same way (cf. bottom row of Fig. \ref{Franz_fig02}).

\begin{figure*}
	\centering
		\includegraphics[width={\textwidth}]{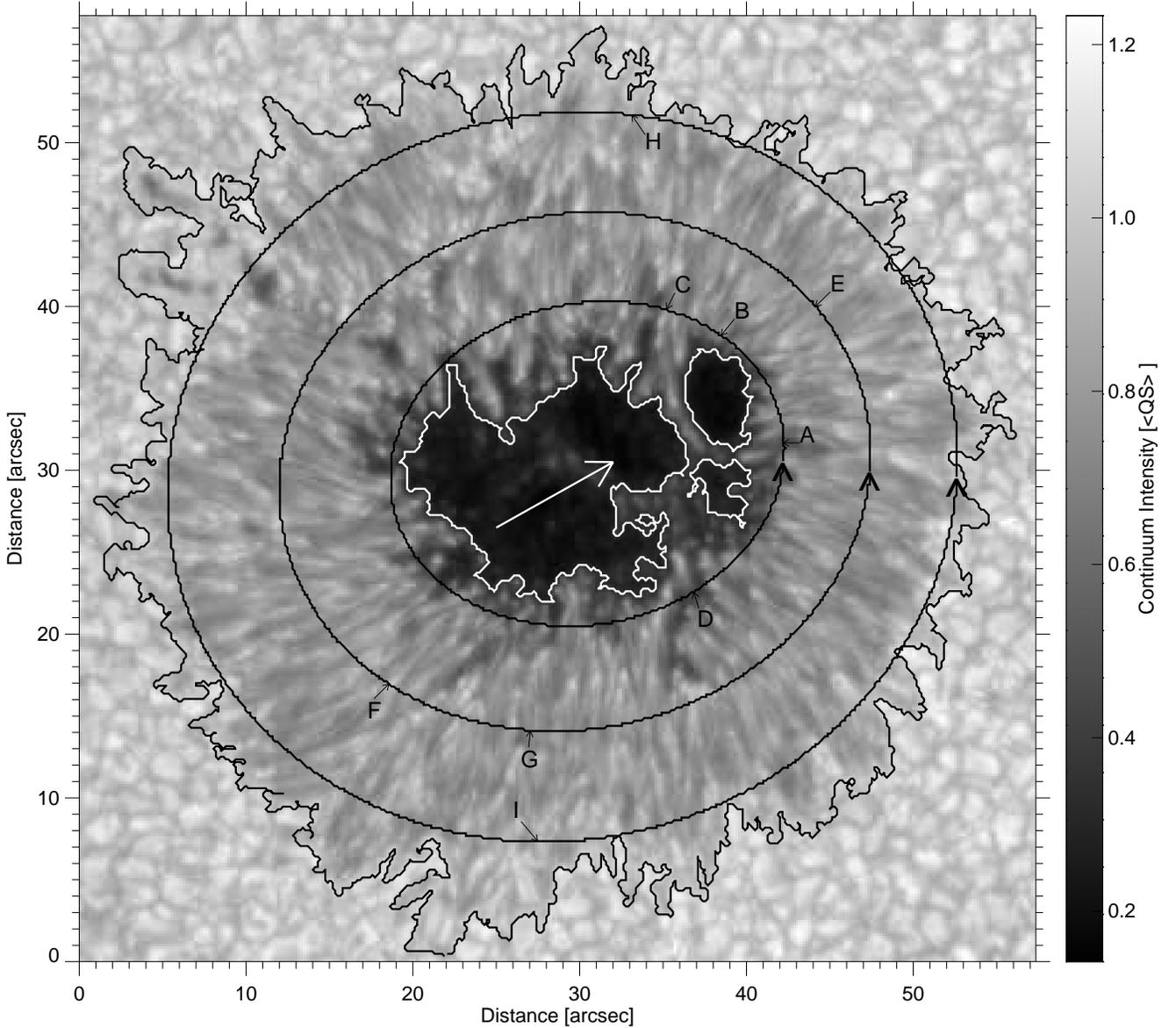}
		\caption{Map of continuum intensity ($\rm{I}_{\rm{c}}$) of Spot D at $\Theta \approx 3^{\circ}$. White/black contours outline the inner/outer penumbral boundary and the arrow points towards disk center. The three ellipses mark tracks in the penumbra at different radii for which $\rm{I}_{\rm{c}}$ and $\rm{v}_{\rm{dop}}$ are investigated in detail in Sect. \ref{morphology}. The path is drawn counterclockwise and starts at the tip of the arrow.}
		\label{Franz_fig03}
\end{figure*}

\begin{figure*}
	\centering
		\includegraphics[width={\textwidth}]{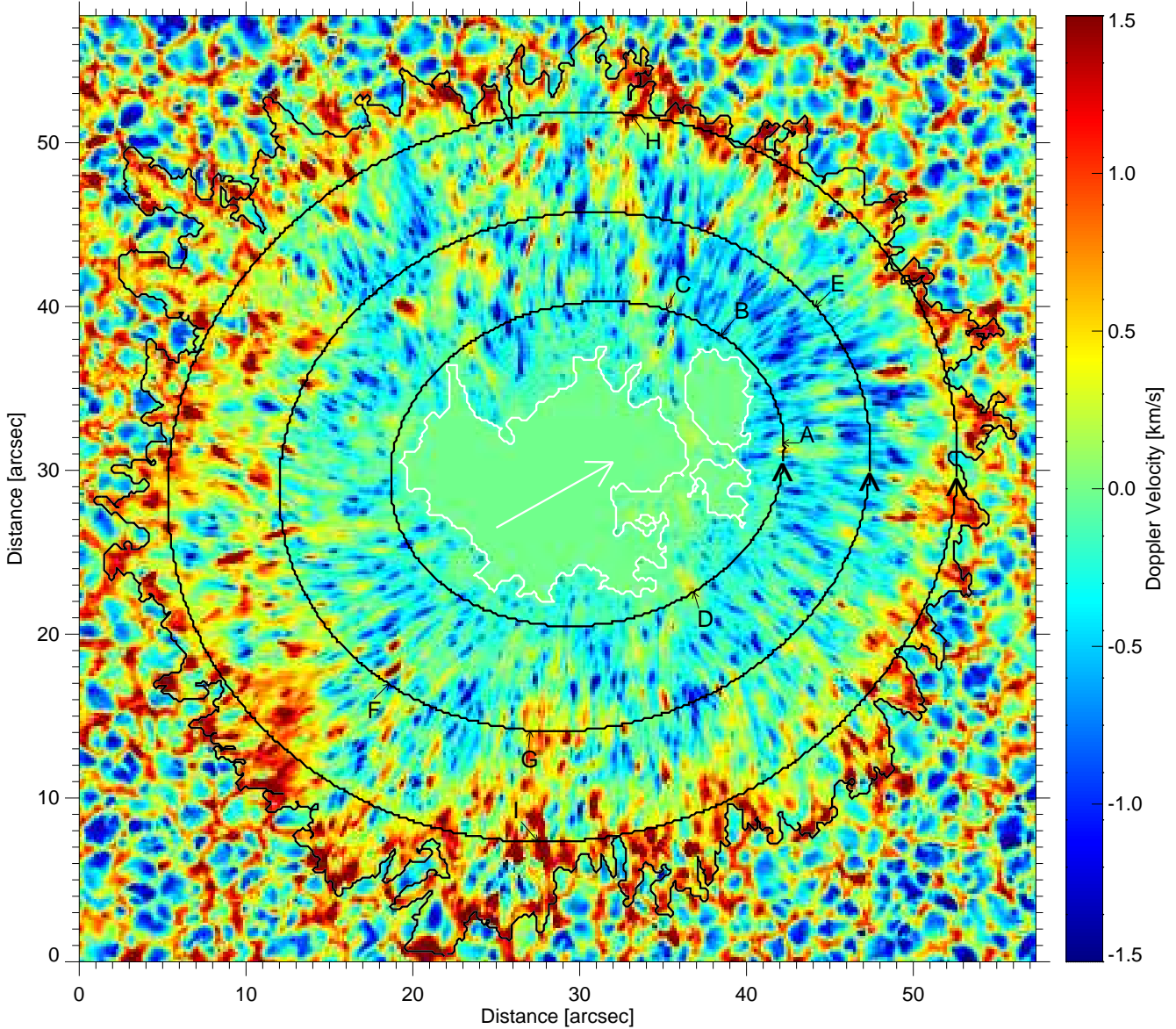}
		\caption{Same as Fig. \ref{Franz_fig03}, but map of LOS velocities in Spot D as derived from the Doppler-shifts in the wing of Fe I 630.15 nm. Up-flows are shown from blue to green, down-flows are drawn from green to red. Note that the map is saturated at $\pm 1.5$ km s$^{-1}$, and the velocities in the umbra are manually set to zero.}
		\label{Franz_fig04}
\end{figure*}

\begin{table}[h]
\begin{center}
	\caption{From left to right: Peak up- and down-flow velocities derived from the Fe I 630.15 line wing, average velocity of penumbra in both lines wings, RMS values of differences between the velocity derived from the two line wings. Negative values correspond to up-, positive values to down-flows.}
	\begin{tabular}{ccccccc}
	\multicolumn{6}{c}{Doppler Velocities in Sunspot Penumbrae \& Quiet Sun Data}\\
		\hline
		\hline
		\\[-2ex]
		{Name} & \multicolumn{2}{c}{v$_{\rm{dop\:630.15}}$ [km s$^{-1}$]} &  \multicolumn{2}{c}{$\langle {\rm{v}_{\rm{dop}} \rangle}$ [m s$^{-1}$]} & \multicolumn{1}{c}{$\Delta\rm{v}_{\rm{dop}}$}\\
		{} & {Up} & {Down} & {\tiny{630.15}} & {\tiny{630.25}} & {RMS [m s$^{-1}$]}\\
		\hline
		{Spot A} & {-2.01} & {8.94} & {157}& {126} & {137}\\
		{Spot B} & {-2.09} & {6.84} & {158}& {127} & {139}\\
		{Spot C} & {-2.24} & {8.88} & {143}& {118} & {168}\\
		{Spot D} & {-2.20} & {5.00} & {105}& {100} & {142}\\
		{QS 1} & {-3.09} & {3.15} & {-159}& {-128} & {83}\\% last 3 columns = line wing values
		{QS 2} & {-3.30} & {3.13} & {-114}& {-104} & {79}\\% last 3 columns = line wing values
		\hline
	\end{tabular}
		\label{Tab_3}
\end{center}
\end{table}

In all sunspot data, the area occupied by up-flows is dominant at small velocities ($|\rm{v}_{\rm{dop}}| < $ 0.4 km s$^{-1}$). The area showing up-flows increases slightly and peaks at $-0.1 \le \rm{v}_{\rm{dop}} \le -0.2 \rm{km\,s}^{-1}$, which seems to be a preferred velocity for up-flows. For larger velocities, this area decreases gradually and is equal in size to the down-flow area at velocities $0.4 \le |\rm{v}_{\rm{dop}}| \le 0.6 \rm{km\,s}^{-1}$, but negligible for $\rm{v}_{\rm{dop}} \le -1.5 \rm{km\,s}^{-1}$. The area showing down-flows, in contrast, decreases monotonously towards larger velocities without any maximum and is the dominant one for $|\rm{v}_{\rm{dop}}| > $ 0.6 km s$^{-1}$. Even at $|\rm{v}_{\rm{dop}}| > 1.5$ km s$^{-1}$, there is a significant area of the penumbra occupied with down-flows.

In the QS data, the area occupied by up-flows is, however, always larger compared to the area showing down-flows, regardless of velocity. Furthermore, the maximum down-flow velocity (3 km s$^{-1}$) is significantly smaller in the QS, while the maximum up-flow velocity ($-$3 km s$^{-1}$) is about 50\% larger as in the penumbra.

These results are consistent with a fact mentioned above: Models and other observations imply an overall blue-shift of the QS, while our investigation yields a significant red-shift of the spatially averaged penumbra.

\subsection{Morphology of Velocity Field}
\label{morphology}

The Doppler map of Spot D shows a concentration of up-flows in the inner penumbra. They often appear as radially elongated patches next to each other separated by areas without any LOS motion. A typical up-flow area is about 2."4 long and 0."5 wide, resulting in a length/width ratio of about 5. Down-flows are predominately visible at the penumbra-QS boundary. They are not elongated, but rather roundish. The length of the down-flow areas ranges from 1."3 to 3."0, while their width extends from 1."3  to 2."0. Thus, with 1 to 1.5 the down-flow length/width ratio is smaller than for the up-flow areas. In order to investigate this morphology quantitatively and to compare the flow field to the continuum intensity, we plotted $\rm{{v}}_{\rm{dop}}$ and $\rm{I}_{\rm{c}}$ along azimuthal paths.

\begin{figure}[h]
	\centering
		\includegraphics[width={\columnwidth}]{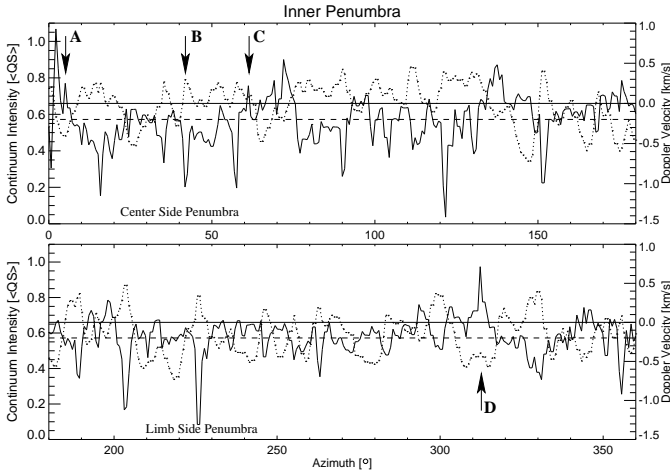}
		\caption{Continuum intensity (dotted) and Doppler velocity (solid) in the penumbra along the inner path shown in Fig. \ref{Franz_fig03} and \ref{Franz_fig04}. Zero velocity is indicated by the solid horizontal line, while the dashed line marks the average velocity along that path.}
		\label{Franz_fig05}
\end{figure}

Fig. \ref{Franz_fig05} shows $\rm{{v}}_{\rm{dop}}$ and $\rm{I}_{\rm{c}}$ in the inner penumbra. What is clearly visible are strong spikes of up-flow with velocities of up to $-$1.5 km s$^{-1}$. Sometimes down-flow peaks are visible, but they are weaker and broader when compared to the up-flows. Areas of up-flow coincide well with areas of increased $\rm{I}_{\rm{c}}$, and areas of down-flow show a reduced $\rm{I}_{\rm{c}}$ (cf. position B \& A in the upper plot of Fig. \ref{Franz_fig05}). This is in agreement with the idea that up-flows transport hot plasma from lower layers to the surface, where it cools radiatively and appears bright. Due to the energy release it becomes more dense than the surrounding plasma and ultimately sinks back below the surface. On average, the plasma shows an up-flow of approx. $-$0.2 km s$^{-1}$ in the inner penumbra.

\begin{figure}[h]
	\centering
		\includegraphics[width={\columnwidth}]{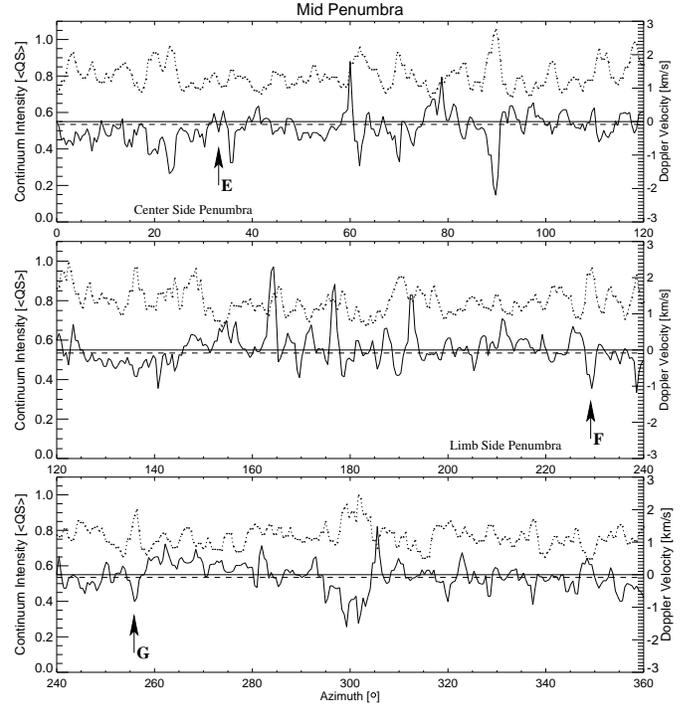}
			\caption{Same as Fig. \ref{Franz_fig05} but along the central path in Fig. \ref{Franz_fig03}.}
		\label{Franz_fig06}
\end{figure}

In the mid penumbra (cf. Fig. \ref{Franz_fig06}), strong up-flow sites are seen less often and down-flows become more prominent. Here the penumbra still shows a net up-flow of about $-$80 m s$^{-1}$, and the correlation between up-flow and enhanced $\rm{I}_{\rm{c}}$ is still valid (cf. F \& G in Fig. \ref{Franz_fig06}). Surprisingly, the strongest up-flow area of Spot D, that is $-2.2$ km s$^{-1}$, is located in the middle penumbra. With 2.3 km s$^{-1}$, the down-flows show velocities of comparable strength.

\begin{figure}[h]
	\centering
		\includegraphics[width={\columnwidth}]{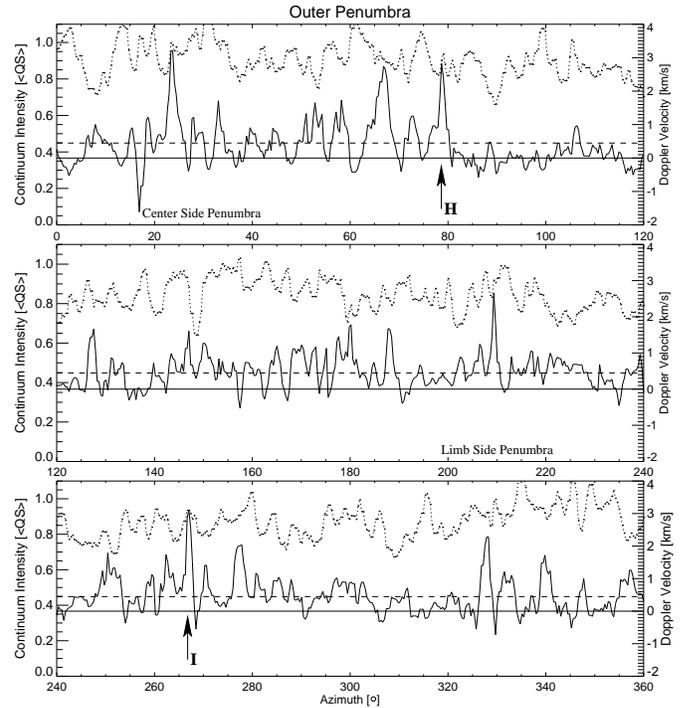}
			\caption{Same as Fig. \ref{Franz_fig05} but for the outer path in Fig. \ref{Franz_fig03}.}
		\label{Franz_fig07}
\end{figure}

In Fig. \ref{Franz_fig07} $\rm{{v}}_{\rm{dop}}$ and $\rm{I}_{\rm{c}}$ of the outer penumbra are plotted. Here strong up-flow patches have almost disappeared and the dominant features are down-flows, with large velocity amplitudes of up to 3.2 km s$^{-1}$. Note that not all up-flows are within the penumbral boundary, but sometimes belong to granulation. Only one up-flow within the penumbra shows velocities of $-$1.6 km s$^{-1}$, while the rest is less strong. On average, the outer penumbra exhibits a down-flow of almost 0.45 km s$^{-1}$. Interestingly, there is no distinct correlation between down-flows and a decrease of $\rm{I}_{\rm{c}}$ in this part of the penumbra anymore. Position H \& I in Fig. \ref{Franz_fig07} indicate two strong down-flow places that show an $\rm{I}_{\rm{c}}$ that is larger compared to the $\rm{I}_{\rm{c}}$ of adjacent pixels which exhibit a weaker down-flow or even an up-flow.

In the inner and mid penumbra, the flow pattern seems to follow the filamentary structure. This is not the case in the outer penumbra anymore, where the down-flow patches appear roundish and show no filamentary structure. In all azimuthal cuts, no distinct difference between the center and the limb side penumbra can be detected in the flow pattern, as it is the case in penumbrae away from disk center \citep{Tritschler2004}.

\subsection{Stokes V signal and magnetized flows}

The Stokes V profile from the down-flow region in, e.g. Fig. \ref{Franz_fig01} shows strong asymmetries and more than two lobes. In order to explain the additional lobes with a simple scenario, we have to assume two magnetic components placed either above or next to each other in the resolution element. In both settings, one of these magnetic components has to be Doppler-shifted (in this example to the red) with respect to the other. Thus, one component would represent a more or less steady magnetized background plasma, while the other component would incorporate strong and magnetized down-flows. 

The first scenario is more likely as it readily explains the shape of the Stokes I profile (cf. Fig. \ref{Franz_fig01}). The line wing is shifted towards the red, while the core is broadened but almost unshifted. Therefore, it is probable that strong velocity gradients along the LOS are present.

Keep in mind that in our case the velocities derived from the line wing of the two iron lines do underestimate the flow velocity. Using the bisector to derive  Doppler velocities will lead to average flow velocity in the resolution element. Assuming two components - one at rest, the other in motion - would automatically increase the velocities of the latter component.

\subsection{Signatures of roll-type convection?}

In the framework of the {\it{gappy}} penumbral model energy is transported by roll-type convection in the penumbral gaps, that is areas without magnetic field below $\tau=1$. Thus, up-flows would be present in the center or the dark core of the filament, while a down-flow should be observable adjacent to it at the position of the lateral brightening \citep{Scharmer2006}. Consequently, a down-up-down flow pattern ought to be visible on a scale of 0.5" in cuts perpendicular to the filamentary structure of penumbrae.

The plots of Fig. \ref{Franz_fig05} to \ref{Franz_fig07}, however, show that strong up- and down-flow patches have a width of more than 0.5" and adjacent to them, the flow does not change its direction, but only eases in strength (cf. position B \& D in Fig. \ref{Franz_fig05}). In rare cases, a reversal of the flow direction is observable on a scale of 0.5" (e.g. position A \& C in Fig. \ref{Franz_fig05}). The depicted flow pattern, however, is just opposite to the signature expected from roll-type convection. One of the very few examples, which shows a down-up-down flow pattern on a scale of less than 1" is indicated as E in Fig. \ref{Franz_fig06} (the peak-to-peak distance of the down-flow patches is 0.6"). A comparison of the maps of $\rm{I}_{\rm{c}}$ and $\rm{{v}}_{\rm{dop}}$, however, shows that this flow pattern does not coincide with a (dark cored) filament. Other down-up-down flow signatures are observable, but have a width larger than 1" (e.g. the down-flow patches at position F show a peak-to-peak distance of 1.4").

A careful analysis of the other data samples does not yield a down-up-down flow pattern on scales of less than 1" that could be interpreted as a signature of roll-type convection in penumbral filaments.
%______________________________________________________________

\section{Summary \& Conclusions}

We presented two independent procedures to calibrate the wavelength scale of {\it{Hinode}} SP data. The first method uses the line core of an average Stokes I profile of the QS. This position was corrected for the CBS of the QS using the results \citep{Beck2005} of model calculations \citep{Borrero2002}. In a second approach, we assumed the umbra at rest and determined the center position, that is midways between the two lobes, of umbral Stokes V profiles with an amplitude asymmetry of less than 1\%. This center position defines a frame of rest on the solar surface \citep{Rezaei2006}. Within the uncertainties, both methods yield the same results.

In order to investigate the flow field in several sunspots close to disk center ($\Theta \le 9^{\circ}$), we determined Doppler velocities in the low photosphere by calculating an average bisector in the wings of both Fe lines. Even though in some cases the amplitudes of the flows derived from Fe I 630.15 and Fe I 630.25 differ significantly, the morphology of the flow field is not affected by these differences. Furthermore, we find several prominent examples of multi-lobed Stokes V profiles which can be explained by a magnetized plasma in combination with strong velocity gradience along the LOS. We therefore conclude that the downward component of the EF in the penumbra is magnetized.

Since all the penumbrae we investigated are located close to disk center, we are able to visualize the LOS motion of the plasma normal to the solar surface, but cannot make a proposition considering the horizontal component of the EF. Other investigations \citep{Rezaei2006}, however, show that at large heliocentric angle, the center position of Stokes V profiles is e.g. blue-shifted on the center-side penumbra, which indicates a magnetized horizontal component of the EF. 

Additionally, we find that the vertical flow pattern in sunspot penumbrae is structured on a small scale. Up-flows appear elongated and predominately in the inner penumbra. Down-flows concentrate at the penumbra-QS boundary and are rather roundish in shape. $\rm{I}_{\rm{c}}$ and $\rm{{v}}_{\rm{dop}}$ along azimuthal cuts in the penumbra show that on average there is an up-flow present in the inner and mid-penumbra, while in the outer penumbra, on average a down-flow is present. The up-flow patches in the inner penumbra and the down-flow locations in the outer penumbra could be interpreted as the sources and the sinks of the EF \citep{Ichimoto2009}, even though an attribution to individual flow channels is challenging \citep{Ichimoto2007}.

Another finding of our investigation is the fact that the properties of the penumbral flow field are strikingly different from that of the QS. In both iron lines, the area of the QS showing up-flows is always larger than the area showing down-flows, regardless of the velocity. This is not surprising as the up-flows are located in the granules and the down-flows are seen in the narrow lanes that form the intergranulum. Interestingly, in the penumbra up-flows dominate only at low velocities $|\rm{{v}}_{\rm{dop}}| \le 0.4$ km s$^{-1}$. At large velocities $|\rm{{v}}_{\rm{dop}}| \ge 0.6$ km s$^{-1}$, the area occupied by down-flows outnumbers the one showing up-flows. This is not only independent of the sunspot sample, but also irrespective of the Fe line we used to derive the Doppler velocity. Furthermore, in all penumbrae up-flows $\rm{{v}}_{\rm{dop}} \le -2$ km s$^{-1}$ seem to be suppressed, while the velocity of up-flows in the QS can reach $\rm{{v}}_{\rm{dop}} = -3$ km s$^{-1}$. On the contrary, down-flows with velocities of 9 km s$^{-1}$ are detected in some penumbrae; three times as large as in the down-flow regions of the QS. Additionally we find that on average the penumbra shows a significant red-shift corresponding to a down-flow of more than 0.1 km s$^{-1}$.

If the  penumbral energy transport is due to convective roll motion, a pronounced down-up-down flow pattern should be observable \citep{Scharmer2006} in azimuthal cuts that cross (dark cored) filaments perpendicularly. In all our data sample, we do not find an indication for a roll-type motion of the plasma on a scale of the width of a filament.

We can speculate why we do not see such a signature, and various explanation are possible. The easiest interpretation of this observation is that no roll-type convection exists in sunspot penumbrae. On the contrary, one could argue that we do not see hints of roll-type convection due to the limited spatial resolution of the {\it{Hinode}} SP. Another explanation could be that the Fe lines are not affected by the convective plasma motion because it is located only slightly above or entirely below $\tau$=1.

The argument of limited spatial resolution always applies. Observations of (dark cored) penumbral filaments show that they have a width of less than 1" sometimes with a dark core around 0.2" in width \citep{Scharmer2002, Suetterlin2004, BellotRubio2005}. In their MHD simulation of an entire sunspot, \citet{Rempel2009} found a down-up-down flow signature with velocity variation of $\pm$1 km s$^{-1}$ across a filament. Even though the observation of the flow morphology in such dark cored penumbral filaments is at the limit of the spatial resolution of the {\it{Hinode}} SP, a distinct down-up-down flow pattern on scales of 0.5" should be observable in cuts perpendicular through the filament. If the plasma motion is entirely below $\tau=1$, we have no chance to ever observe it. Yet, any plasma motion below $\tau=1$ will affect the higher atmospheric layers as well. So even if there is convective motion only below the surface, we should see at least a weak Doppler signal in the far wings of the Fe lines.

A more detailed investigation of the flow field in small scale features of the penumbra, e.g. bright penumbral grains, filaments, etc., will be addressed in a forthcoming publication using the same data. Furthermore, it will be interesting to compare synthetic Stokes profiles computed from realistic simulations of radiative magneto-convection with our observations.

\begin{acknowledgements}
We want to thank O. Steiner and R. Rezaei for fruitful discussions, and W. Schmidt for his valuable comments on the manuscript. Part of this work was supported by the \emph{Deut\-sche For\-schungs\-ge\-mein\-schaft, DFG\/} project number Schl.~514/3-1. {\it{Hinode}} is a Japanese mission developed and launched by ISAS/JAXA, with NAOJ as domestic partner and NASA and STFC (UK) as international partners. It is operated by these agencies in cooperation with ESA and NSC (Norway).
\end{acknowledgements}
\bibliographystyle{aa}
\bibliography{Franz}

\begin{thebibliography}{30}
\expandafter\ifx\csname natexlab\endcsname\relax\def\natexlab#1{#1}\fi

\bibitem[{{Auer} \& {Heasley}(1978)}]{Auer1978}
{Auer}, L.~H. \& {Heasley}, J.~N. 1978, \aap, 64, 67

\bibitem[{Beck(2005)}]{Beck2005}
Beck, C. 2005, PhD thesis, The University of Freiburg

\bibitem[{{Beckers}(1977)}]{Beckers1977}
{Beckers}, J.~M. 1977, \apj, 213, 900

\bibitem[{{Bellot Rubio}(2009)}]{BellotRubio2009}
{Bellot Rubio}, L.~R. 2009, ArXiv e-prints : 0903.3619

\bibitem[{{Bellot Rubio} {et~al.}(2005){Bellot Rubio}, {Langhans}, \&
  {Schlichenmaier}}]{BellotRubio2005}
{Bellot Rubio}, L.~R., {Langhans}, K., \& {Schlichenmaier}, R. 2005, \aap, 443,
  L7

\bibitem[{{Bellot Rubio} {et~al.}(2007){Bellot Rubio}, {Tsuneta}, {Ichimoto},
  {Katsukawa}, {Lites}, {Nagata}, {Shimizu}, {Shine}, {Suematsu}, {Tarbell},
  {Title}, \& {del Toro Iniesta}}]{BellotRubio2007}
{Bellot Rubio}, L.~R., {Tsuneta}, S., {Ichimoto}, K., {et~al.} 2007, \apjl,
  668, L91

\bibitem[{Borrero \& Bellot~Rubio(2002)}]{Borrero2002}
Borrero, J.~M. \& Bellot~Rubio, L.~R. 2002, \aap, 385, 1056

\bibitem[{{Centeno} {et~al.}(2009){Centeno}, {Lites}, {de Wijn}, \&
  {Elmore}}]{Centeno2009}
{Centeno}, R., {Lites}, B.~W., {de Wijn}, A.~G., \& {Elmore}, D. 2009, ArXiv
  e-prints : 0902.0027

\bibitem[{{Dravins}(1982)}]{Dravins1982}
{Dravins}, D. 1982, \araa, 20, 61

\bibitem[{{Heinemann} {et~al.}(2007){Heinemann}, {Nordlund}, {Scharmer}, \&
  {Spruit}}]{Heinemann2007}
{Heinemann}, T., {Nordlund}, {\AA}., {Scharmer}, G.~B., \& {Spruit}, H.~C.
  2007, \apj, 669, 1390

\bibitem[{{Ichimoto} {et~al.}(2007){Ichimoto}, {Shine}, {Lites}, {Kubo},
  {Shimizu}, {Suematsu}, {Tsuneta}, {Katsukawa}, {Tarbell}, {Title}, {Nagata},
  {Yokoyama}, \& {Shimojo}}]{Ichimoto2007}
{Ichimoto}, K., {Shine}, R.~A., {Lites}, B., {et~al.} 2007, \pasj, 59, 593

\bibitem[{{Ichimoto} \& {SOT/Hinode-team}(2009)}]{Ichimoto2009}
{Ichimoto}, K. \& {SOT/Hinode-team}. 2009, ArXiv e-prints

\bibitem[{{Khomenko} \& {Collados}(2007)}]{Khomenko2007}
{Khomenko}, E. \& {Collados}, M. 2007, \apj, 659, 1726

\bibitem[{Lites {et~al.}(2001)Lites, Elmore, \& Streander}]{Lites2001}
Lites, B.~W., Elmore, D.~F., \& Streander, K.~V. 2001, in ASP Conf. Ser. 236
  Advanced Solar Polarimetry, ed. M. Sigwarth (San Francisco: ASP), 33

\bibitem[{Rempel {et~al.}(2009)Rempel, Sch\"ussler, \& Kn\"olker}]{Rempel2009}
Rempel, M., Sch\"ussler, M., \& Kn\"olker, M. 2009, \apj, 691, 640

\bibitem[{{Rezaei} {et~al.}(2006){Rezaei}, {Schlichenmaier}, {Beck}, \& {Bellot
  Rubio}}]{Rezaei2006}
{Rezaei}, R., {Schlichenmaier}, R., {Beck}, C., \& {Bellot Rubio}, L.~R. 2006,
  \aap, 454, 975

\bibitem[{Ruiz~Cobo \& Bellot~Rubio(2008)}]{RuizCobo2008}
Ruiz~Cobo, B. \& Bellot~Rubio, L. 2008, \aap, 488, 749

\bibitem[{Scharmer {et~al.}(2002)Scharmer, Gudiksen, Kiselman, L\"ofdahl, \&
  Rouppe van~der Voort}]{Scharmer2002}
Scharmer, G.~B., Gudiksen, B.~V., Kiselman, D., L\"ofdahl, M.~G., \& Rouppe
  van~der Voort, L. H.~M. 2002, Nature, 420, 151

\bibitem[{{Scharmer} {et~al.}(2008){Scharmer}, {Narayan}, {Hillberg}, {de la
  Cruz Rodriguez}, {L{\"o}fdahl}, {Kiselman}, {S{\"u}tterlin}, {van Noort}, \&
  {Lagg}}]{Scharmer2008}
{Scharmer}, G.~B., {Narayan}, G., {Hillberg}, T., {et~al.} 2008, \apjl, 689,
  L69

\bibitem[{{Scharmer} \& {Spruit}(2006)}]{Scharmer2006}
{Scharmer}, G.~B. \& {Spruit}, H.~C. 2006, \aap, 460, 605

\bibitem[{Schlichenmaier {et~al.}(1998{\natexlab{a}})Schlichenmaier, Jahn, \&
  Schmidt}]{Schlichenmaier1998a}
Schlichenmaier, R., Jahn, K., \& Schmidt, H.~U. 1998{\natexlab{a}}, \apj, 493L,
  121

\bibitem[{Schlichenmaier {et~al.}(1998{\natexlab{b}})Schlichenmaier, Jahn, \&
  Schmidt}]{Schlichenmaier1998b}
Schlichenmaier, R., Jahn, K., \& Schmidt, H.~U. 1998{\natexlab{b}}, \aap, 337,
  897

\bibitem[{{Sch{\"u}ssler} \& {V{\"o}gler}(2006)}]{Schuessler2006}
{Sch{\"u}ssler}, M. \& {V{\"o}gler}, A. 2006, \apjl, 641, L73

\bibitem[{{Solanki}(1993)}]{Solanki1993b}
{Solanki}, S.~K. 1993, Space Science Reviews, 63, 1

\bibitem[{Solanki \& Montavon(1993)}]{Solanki1993a}
Solanki, S.~K. \& Montavon, C. A.~P. 1993, \aap, 275, 283

\bibitem[{Spruit \& Scharmer(2006)}]{Spruit2006}
Spruit, H.~C. \& Scharmer, G.~B. 2006, \aap, 447, 343

\bibitem[{{S{\"u}tterlin} {et~al.}(2004){S{\"u}tterlin}, {Bellot Rubio}, \&
  {Schlichenmaier}}]{Suetterlin2004}
{S{\"u}tterlin}, P., {Bellot Rubio}, L.~R., \& {Schlichenmaier}, R. 2004, \aap,
  424, 1049

\bibitem[{{Tritschler} {et~al.}(2004){Tritschler}, {Schlichenmaier}, {Bellot
  Rubio}, \& {the KAOS Team}}]{Tritschler2004}
{Tritschler}, A., {Schlichenmaier}, R., {Bellot Rubio}, L.~R., \& {the KAOS
  Team}. 2004, \aap, 415, 717

\bibitem[{{Tsuneta} {et~al.}(2008){Tsuneta}, {Ichimoto}, {Katsukawa}, {Nagata},
  {Otsubo}, {Shimizu}, {Suematsu}, {Nakagiri}, {Noguchi}, {Tarbell}, {Title},
  {Shine}, {Rosenberg}, {Hoffmann}, {Jurcevich}, {Kushner}, {Levay}, {Lites},
  {Elmore}, {Matsushita}, {Kawaguchi}, {Saito}, {Mikami}, {Hill}, \&
  {Owens}}]{Tsuneta2008}
{Tsuneta}, S., {Ichimoto}, K., {Katsukawa}, Y., {et~al.} 2008, \solphys, 249,
  167

\bibitem[{{Vargas Dom{\'{\i}}nguez} {et~al.}(2008){Vargas Dom{\'{\i}}nguez},
  {Rouppe van der Voort}, {Bonet}, {Mart{\'{\i}}nez Pillet}, {Van Noort}, \&
  {Katsukawa}}]{Vargas2008}
{Vargas Dom{\'{\i}}nguez}, S., {Rouppe van der Voort}, L., {Bonet}, J.~A.,
  {et~al.} 2008, \apj, 679, 900

\end{thebibliography}
\end{document}